\documentclass{article}

\usepackage{arxiv}

\usepackage[utf8]{inputenc} 
\usepackage[T1]{fontenc}    
\usepackage{hyperref}       
\usepackage{url}            
\usepackage{booktabs}       
\usepackage{amsfonts}       
\usepackage{nicefrac}       
\usepackage{microtype}      
\usepackage{lipsum}
\usepackage{float}
\usepackage{subcaption}
\usepackage{graphicx}
\usepackage{amsmath,amssymb,amsfonts}
\graphicspath{ {./images/} }

\title{Generating Seamless Virtual Immunohistochemical Whole Slide Images with Content and Color Consistency}

\author{
 Sitong Liu \\
  University of Washington\\
  \texttt{sitonl2@uw.edu} \\
   \And
 Kechun Liu \\
  University of Washington\\
  \texttt{kechun@cs.washington.edu } \\
  \And
 Samuel Margolis \\
  University of California, Los Angeles\\
  \texttt{SMargolis@mednet.ucla.edu} \\
  \AND
  Wenjun Wu \\
  University of Washington \\
  \texttt{wenjunw@uw.edu} \\
  \And
  Stevan R. Knezevich \\
  \texttt{stevanrk@gmail.com} \\
  \And
  David E Elder \\
  University of Pennsylvania \\
  \texttt{David.Elder@pennmedicine.upenn.edu} \\
   \And
  Megan M. Eguchi \\
  University of Pennsylvania \\
  \texttt{MEguchi@mednet.ucla.edu} \\
  \And
  Joann G Elmore \\
  University of California, Los Angeles \\
  \texttt{JElmore@mednet.ucla.edu} \\
  \And
  Linda Shapiro \\
  University of Washington \\
  \texttt{shapiro@cs.washington.edu} \\
}

\begin{document}
\maketitle
\begin{abstract}
Immunohistochemical (IHC) stains play a vital role in a pathologist's analysis of medical images, providing crucial diagnostic information for various diseases. Virtual staining from hematoxylin and eosin (H\&E)-stained whole slide images (WSIs) allows the automatic production of other useful IHC stains without the expensive physical staining process. However, current virtual WSI generation methods based on tile-wise processing often suffer from inconsistencies in content, texture, and color at tile boundaries. These inconsistencies lead to artifacts that compromise image quality and potentially hinder accurate clinical assessment and diagnoses. To address this limitation, we propose a novel consistent WSI synthesis network, \textit{CC-WSI-Net}, that extends GAN models to produce seamless synthetic whole  slide images. Our \textit{CC-WSI-Net} integrates a content- and color-consistency supervisor, ensuring consistency across tiles and facilitating the generation of seamless synthetic WSIs while ensuring Sox10 immunohistochemistry accuracy in melanocyte detection. We validate our method through extensive image-quality analyses, objective detection assessments, and a subjective survey with pathologists. By generating high-quality synthetic WSIs, our method opens doors for advanced virtual staining techniques with broader applications in research and clinical care.
\end{abstract}


\section{Introduction}
Digital whole slide images (WSIs) play a crucial role in histopathological studies. Compared to traditional diagnosis of a glass slide using a microscope, a WSI not only simulates the functionality of traditional microscopes \cite{elmore2017randomized} but also offers multiple additional benefits, including side-by-side comparisons, remote accessibility, and convenience. \cite{al2012digital}. In addition, the WSI can be useful for diagnosing certain diseases, such as hepatocellular carcinoma \cite{melo2020whole}, lymphoma \cite{amin2019validation}, breast cancer, and melanoma. To improve diagnostic accuracy, tissues are stained to enhance the visibility of cellular components in WSIs\cite{ojukwu2024immunohistochemistry}. Hematoxylin and eosin (H\&E) is the most prevalent staining technique, where hematoxylin colors cell nuclei a purplish-blue and eosin stains the extracellular matrix and cytoplasm pink. However, H\&E-stained WSIs often provide only a general overview, displaying different cell types with similar coloration, which can be insufficient for diagnosing certain diseases and may lead to detection failures. Immunohistochemical (IHC) stains are therefore widely used in addition to the H\&E for making a clinical diagnosis. IHC staining employs monoclonal and polyclonal antibodies to highlight specific tissue components by identifying the presence and location of specific antigens within tissue samples\cite{duraiyan2012applications}.  IHC stains have wide applications in medical diagnoses, providing critical insights for accurate disease detection and characterization. This technique is invaluable for identifying specific tumor antigens, offering more precise and detailed insights compared to H\&E staining. For instance, TTF1 is a helpful marker for pulmonary adenocarcinoma, while the p40 antibody is a helpful marker for squamous cell carcinoma. \cite{yatabe2019best}; the Sox10 antibody is used in specialized IHC staining to highlight melanocytes for diagnosing melanoma. 

\begin{figure}[ht]
    \centering
    \includegraphics[height=7cm]{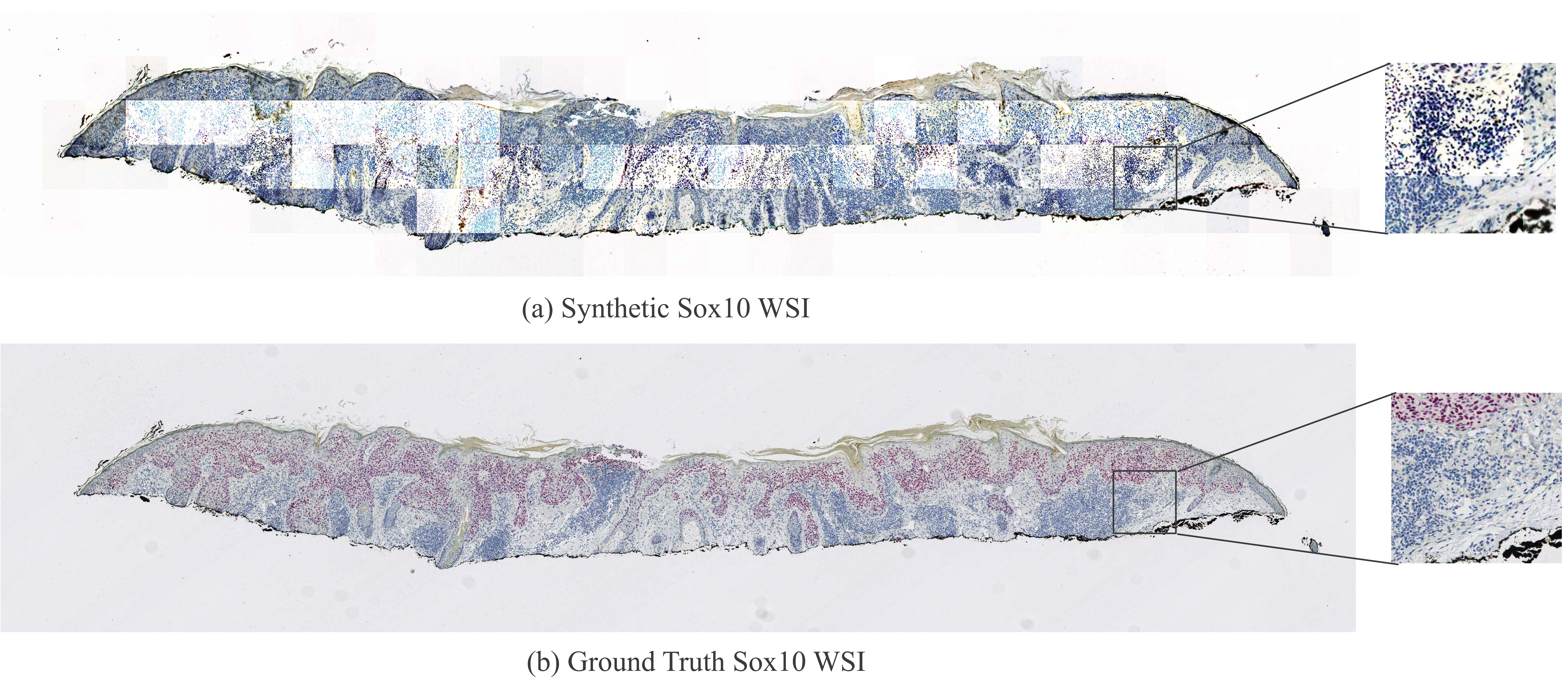} 
    \caption{Visualization of the color and content inconsistency problem. (a) shows a Sox10 WSI generated by the VSGD-Net, VSGD-Net, while (b) displays the corresponding ground truth Sox10 WSI. The zoomed-in crop of (a) reveals inconsistencies in content and color compared to the zoomed-in crop of (b).}
    \label{fig:inconsistency}
\end{figure}

Even though adding the IHC staining to H\&E information offers improved diagnostic capabilities compared to H\&E staining alone, the process  of obtaining IHC staining can result in longer processing times and higher costs. Consequently, IHC staining is not routinely used as frequently as H\&E staining. With advancements in computer vision and machine learning, image synthesis has become a crucial research area in pathology. This technique allows the generation of synthetic WSIs with different types of stainings from the original H\&E WSIs. Virtual staining, which can be faster and less expensive, provides a viable alternative to obtaining real IHC stains for use in analyses.

Several studies have proposed methods for virtual staining. Liu \textit{et al.} developed VSGD, which not only virtually restains H\&E patches to Sox10 patches but also detects melanocytes within those H\&E patches \cite{liu2023vsgd}. Haan \textit{et al.} utilized GANs to virtually stain H\&E patches in WSIs with special stains, such as PAS, MT, and JMS \cite{de2021deep}. While virtual staining is emerging as a promising field with significant potential to enhance disease diagnosis procedures, these methods have limitations. Most studies are constrained to patch-level analysis due to the gigapixel scale of WSIs. While WSIs offer substantial diagnostic benefits with their high resolution, often exceeding $10,000 \times 10,000$ pixels, the high dimensionality of such data presents challenges in fitting an entire WSI into memory. Given computational resource limitations, most research to date has segmented WSIs into smaller patches for analysis. However, analyzing and handling WSIs in patches has significant caveats from a clinical perspective: important details, such as structural features and morphological information, can be lost, making precise assessment more challenging or even impossible. These missing details are critical for accurate disease diagnosis.

Analyzing virtually stained images at the whole-slide level is crucial for diagnosis. This allows pathologists to assess the entire tissue context, which is essential for accurate evaluation. However, a significant challenge arises when we attempt to achieve this by simply stitching synthesized patches back into a WSI. As shown in Fig.~\ref{fig:inconsistency}, stitching often introduces artifacts, leading to inconsistencies in content, texture, and color between adjacent patches. This significantly reduces the quality of the final synthetic WSI and potentially hinders its diagnostic utility. \par
\textbf{Bridging the gap between patch-level analysis and slide-level analysis is essential for improving diagnostic quality and effectiveness.} Several projects have addressed this challenge by focusing on generating high resolution virtually stained WSIs directly. For instance, Harb \textit{et al.} introduced a diffusion-based method for creating synthetic histopathological WSIs. Lahiani \textit{et al.} proposed a novel Perceptual Embedding Consistency (PEC) loss to to improve color consistency within stitched patches. However, a crucial question remains unanswered: can these virtual stains, while achieving visually compelling virtually stained WSIs, provide diagnostically relevant information that can ultimately improves diagnostic accuracy.

To address the limitations of existing methods and ensure virtual staining is not only natural-looking, but also accurate in detecting diagnostically relevant structures,  we introduce a novel consistent WSI synthesis network, \textit{CC-WSI-Net}, that builds upon VSGD-Net\cite{liu2023vsgd}. Our method incorporates a content and color consistency module, enabling the generation of seamless WSIs without compromising stain accuracy. Extensive image quality analysis and accuracy assessments validate the superiority of our method. Furthermore, we conducted a subjective survey among three experienced pathologists to evaluate the clinical quality of the synthetic images. Our main contributions are summarized as follows:
\begin{enumerate}
    \item We developed a framework, CC-WSI-Net, that can be flexbly integrated into existing patch-based GAN models, facilitating the synthesis of WSIs free from color and content inconsistency issues.
    \item CC-WSI-Net is the first model to produce virtually stained WSIs that pathologists confirm are both realistic in appearance and diagnostically equivalent to traditional Sox10 staining.
\end{enumerate}

\section{Related Work}
\label{sec:Related Work}
\subsection{Image Generation}
Machine learning advancements in computer vision have spurred the development of numerous image generation models. Among these, Generative Adversarial Networks (GANs) are particularly notable. Building upon the successful adversarial learning, numerous GAN variants have been developed for image synthesis. Karras \textit{et al.} proposed StyleGAN, which provides further control over the style of the generated images using Adaptive Instance Normalization (AdaIN) within the generator\cite{karras2019style}. However, AdaIN can sometimes lead to water drop artifacts. To address this issue, Karras \textit{et al.} introduced StyleGAN2, which replaces AdaIN with weight demodulation in each layer \cite{karras2020analyzing}. This not only eliminates artifacts but also allows for more precise style control. To exert greater control over image content, researchers developed conditional GANs (cGANs). Pix2pixHD is one of the most popular cGANs for high-resolution image-to-image translation tasks \cite{wang2018high}. Leveraging StyleGAN2 and cGAN, Afifi \textit{et al.} designed reHistoGAN, a model that recolors an input image with the target color scheme using the 2-D histogram of the target image \cite{afifi2021histogan}.

Although GANs have demonstrated remarkable capabilities in manipulating image content, their training process often involves complex architectures and paired datasets. Diffusion models, on the other hand, use a more iterative approach to image generation, starting with random noise and progressively refining it into a realistic image. In recent years, diffusion models have significantly advanced the research in generative models. For example, Ho \textit{et al.} designed the Denoising Diffusion Probabilistic Model (DDPM), which consists of two main processes: forward and backward. The forward diffusion process incrementally adds small amounts of Gaussian noise to input images over multiple steps, gradually transforming the data into pure Gaussian noise. The backward process iteratively removes the noise and reconstructs the original image from the pure noise \cite{ho2020denoising}. Several diffusion models have been proposed for image-to-image translation. Leveraging the Brownian Bridge concept, Li \textit{et al.} introduced the Brownian Bridge Diffusion Model  (BBDM) framework \cite{li2023bbdm}. Unlike DDPM which reconstructs the original image, BBDM ends with the target image at the final Step. BBDM leverages a VQGAN to encode and decode the source and target images, and the noise is added and removed within latent space during the diffusion process.

\subsection{Virtual Staining} 
The use of deep learning models in image generation has significantly boosted medical imaging research, especially in virtual staining. Given the potential of virtual staining of WSIs to support disease diagnosis, numerous studies have utilized deep learning models for these purposes. Among these models, CycleGAN is one of the most popular for virtual staining due to its ability to work without requiring paired training data. Xu \textit{et al.} proposed a conditional CycleGAN (cCGAN) network designed to transform H\&E stained images into IHC stained images \cite{xu2019gan}. Mercan \textit{et al.} employed a GAN and a CycleGAN to virtually stain breast cancer tissue from H\&E to PHH3 and vice versa \cite{9098409}. Liu \textit{et al.} developed a virtual staining method using a CycleGAN to translate neuroendocrine tumor H\&E stained images into Ki-67 stained images \cite{liu2021generation}. In addition, Anonymous \textit{et al.} leverages a paired H\&E plus IHC dataset and extends Pix2PixHD \cite{wang2018high} to build VSGD-Net that learns H\&E to IHC staining via the melanocyte detection task.

\subsection{Breaking the size limits: WSI Synthesis}
Although the virtual staining methods achieve great success in translating image modality, these approaches typically operate only at the patch level. Unfortunately, the small synthetic patches are not as helpful to pathologists in making their diagnoses as WSIs that provide the complete picture. In order to break the size limits and generate Whole Slide Images (WSIs), researchers have explored various solutions via different architectures. For instance, Harb \textit{et al.} introduced a diffusion-based method for creating synthetic histopathological WSIs \cite{oyelade2024saltgan}. It begins with generating a low-resolution image from noise using a diffusion-based model. This initial low-resolution image serves as the foundation for a subsequent upscaling process, which is enhanced incrementally through a structured coarse-to-fine sampling scheme. Each stage of this process not only increases the image's resolution but also integrates increasingly finer details. However, this model is limited to synthesizing WSIs at resolutions up to $65,536 \times 65,536$ pixels. Lahiani \textit{et al.} proposed a novel Perceptual Embedding Consistency (PEC) loss to enhance CycleGAN, aiming to learn color, contrast, and brightness features to achieve color consistency among patches \cite{lahiani2020seamless}. Sun \textit{et al.} proposed a Bi-directional Feature Fusion Generative Adversarial Network (BFF-GAN) that integrates global and local features through a dual-branch architecture. The global branch processes down-sampled whole images, while the local branch handles high-resolution patches. Feature fusion occurs bi-directionally, with patch-wise attention enhancing feature expression, effectively addressing differences in color and brightness between adjacent patches \cite{sun2023bi}. Although these studies have demonstrated satisfying performance in synthesizing WSIs with different stainings without inconsistency problems, they do not account for the clinical quality and effectiveness of the stain in their evaluations. This consideration is vital for the application of these methods to clinical practice and disease diagnosis.

\section{Methodology}
\subsection{Problem Setup}
Given the potential significance of accurate virtually stained WSIs in improving diagnostic accuracy and clinical care, we aim to study the synthesis of seamless WSIs that may be clinically effective. To the best of our knowledge, VSGD-Net \cite{liu2023vsgd} represents the state-of-the-art method for synthesizing realistic IHC stained patches from H\&E stained images while simultaneously detecting diagnostically relevant regions. Hence, we base our study on VSGD-Net, focusing on extending it to synthesize WSIs while ensuring the stain accuracy and effectiveness to support clinical use.

VSGD-Net is built upon Pix2PixHD, with an additional detection network that utilizes the intermediate features in the generator to identify certain objects. It was first studied and applied to a melanoma biopsy dataset, in which the diagnosis is formed based on the assessment of melanocytes' growth patterns and location distributions. When assessing skin biopsies for the presence of possible invasive melanoma, the routine H\&E staining is often insufficient to highlight melanocytes while the costly Sox10 IHC stain can point out melanocytes and thus aid in diagnosis. VSGD-Net is particularly effective in such a scenario, because the detection of melanocytes and the virtual staining from H\&E to Sox10 can boost each other, thus achieving state-of-the-art performance in both melanocyte detection and image synthesis tasks. VSGD-Net caused excitement among pathologists who hoped to use it for generating synthetic stained images for use in their research and clinical practices. Thus our work was focused on solving the color and content inconsistency problems that kept it from working on a whole slide image instead of just small patches.

\begin{figure}
    \centering
    \includegraphics[height=6cm]{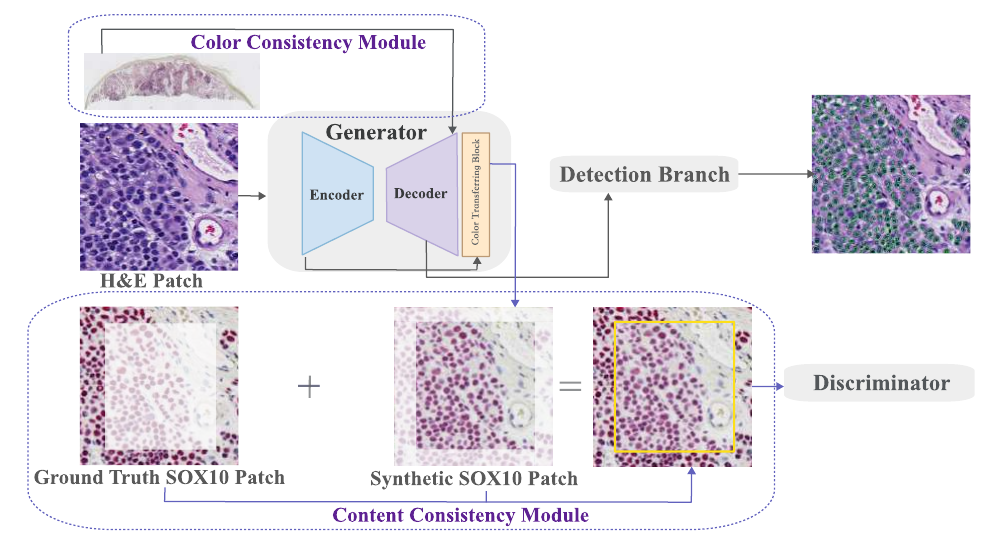} 
    \caption{Overview of \textit{CC-WSI-Net}: A content consistency module and a color consistency module are flexibly applied to the original VSGD-Net architecture, which includes a generator, a discriminator and a detection  branch.}
    \label{framework overview}
\end{figure}

Fig. \ref{framework overview} presents an overview of our \textit{CC-WSI-Net}, which integrates a content consistency module and a color consistency module into the original VSGD-Net. These novel modules provide effective control over the synthesized patch images, successfully resolving stitching artifacts. Detailed explanations are provided in the following subsections.

\subsection{Content Consistency}
\begin{figure}
    \centering
    \begin{subfigure}[b]{0.35\textwidth} 
        \centering
        \includegraphics[width=\textwidth]{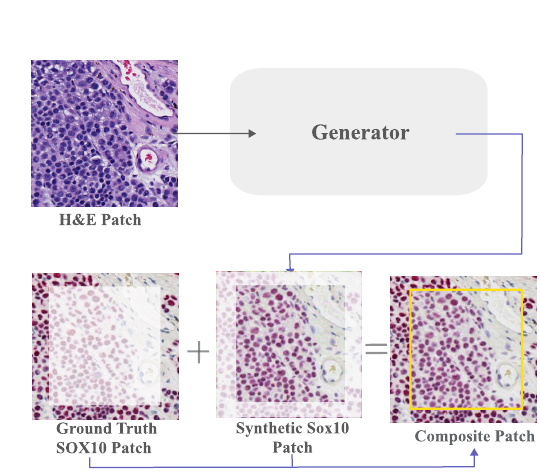}
        \caption{Content consistency module}
        \label{Content consistency module}
    \end{subfigure}
    \hspace{\fill} 
    \begin{subfigure}[b]{0.6\textwidth} 
        \centering
        \includegraphics[width=\textwidth]{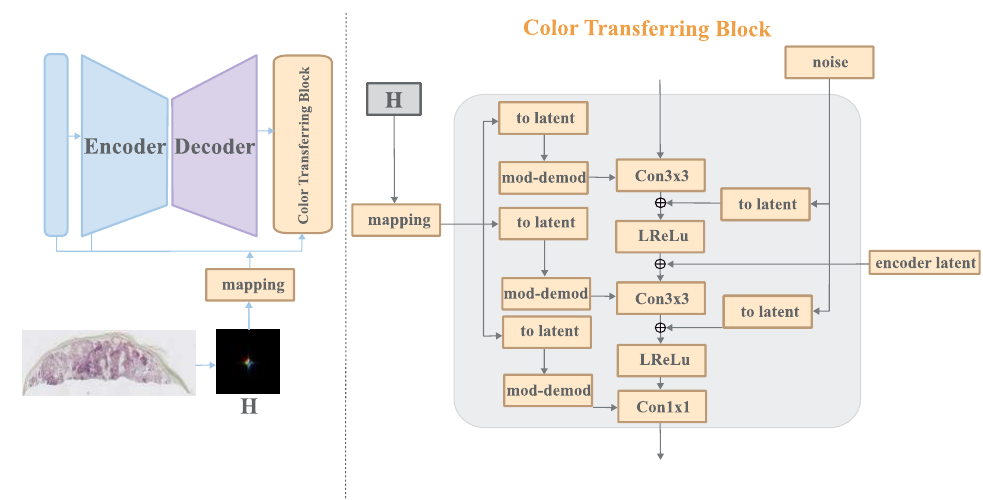}
        \caption{Color consistency module}
        \label{Color consistency module}
    \end{subfigure}
    \caption{(a) provides the details of the content consistency module. The left side of (b) illustrates the overall generator with the color consistency module, while the right side shows the architecture of the color transferring block. \textit{H} represents the 2-D color histogram of the WSI used as the color condition, and \textit{mapping} refers to a fully connected network.}
    \label{Framework details}
\end{figure}

To enforce content consistency over the borders of patches, we leverage the discriminator to distinguish a composite image. As shown in Fig. \ref{Content consistency module}, during training, we only keep the center crop of the synthesized image and replace the surrounding regions with data from the ground truth image to create the composite image. This approach ensures that the center crop images are learned with additional contextual information from the whole patch, leading to improved consistency and accuracy in the final output. We then input the composite image to the discriminator to assess whether this image is realistic. This design not only verifies the overall realism of the image but also allows the discriminator to ensure that the synthesized central content blends smoothly with the genuine surrounding border. 
During inference, only the center crops are utilized to generate the WSI, and the surrounding regions are discarded.

\subsection{Color Consistency}
To address the color inconsistency problem, we propose a color consistency module which can be flexibly applied to the generator. 
We replace the last convolution layer in the decoder with a color transferring block. This module is borrowed from the recolor head in ReHistoGAN \cite{afifi2021histogan}. As shown in Fig. 3(b), the color transferring block has three inputs: noise sampled from a standard normal distribution,  the encoder features, and the color histogram condition. To preserve the fine structure details in the synthesized images, the first two layers in the encoder are passed to the color transferring block. In addition, the color transferring block takes a 2-D color histogram from a 2.5x magnification WSI as a color condition to guide the color distribution of the generated image.

This modification of the generator aims to consider both input image structure information and the target color histogram in the recoloring process. In the color transferring block, the color is further revised based on the target histogram. For each patch in the same WSI, a color histogram will be employed to ensure the same color condition. In our experiments, we chose the 2-D histogram of the whole Sox10-stained WSI as the color condition, because patch-level color histograms are not available during inference, and the histograms for individual patches exhibit distinct differences due to the various quantities of melanocytes. Thus, we leverage the 2-D histogram of an entire WSI as the color condition during the training process, providing weak supervision to guide the synthesis. During inference time, a WSI histogram is sampled to condition the virtual staining.

\subsection{Training Process}
Our base model, VSGD-Net, was optimized by GAN Loss, Similarity Loss \(L_{feat}\) and Detection Loss \(L_{DET}\), which are combined as:
\begin{equation}
min_{G} ( \max_{D} \sum_{i=1}^2 \left( \log(D_i(X_s)) + \log(1 - D_i(G(X_h))) \right) + \lambda  * L_{\text{feat}} + L_{\text{DET}} ),
\end{equation}
where \(G\) is the generator, \(D_{i}\) is the Discriminator, and \(X_s\) and \(X_h\) are a pair of Sox10 and H\&E images. $L_{\text{feat}}$ is the feature matching loss in discriminator layers, while $L_{\text{DET}}$ consists of the losses of classification, localization and segmentation in the detection branch\cite{liu2023vsgd}.

Analogous to the approach employed in HistoGAN, we utilize the Color-matching Histogram Loss to supervise the color scheme of the synthetic Sox10 image. The histogram loss is given as:
\begin{equation}
C(\mathbf{H}_g, \mathbf{H}_s) = \frac{1}{2} \left\| \mathbf{H}_g^{1/2} - \mathbf{H}_s^{1/2} \right\|_2,
\end{equation}
where \(\mathbf{H}_g\) and \(\mathbf{H}_s\) are the histograms of the ground truth Sox10 patch and synthetic Sox10 patch respectively.

We observe that each slice exhibits a distinct color scheme. Notably, the histogram loss appears to be predominantly influenced by variations in background color, resulting in a greater backpropagation of loss from background patches than from tissue patches. To learn more effectively from the tissue patches, we introduced an adaptive weighting mechanism to each histogram loss, proportional to the tissue content in each patch. Consequently, patches with a higher proportion of tissue are penalized more heavily, while those with greater amounts of background receive a lesser penalty. The equation for adaptive weight is defined as follow:
\begin{equation}
\text{W}_{h} = sigmoid(\text{tissue portion})
\label{eq1}
\end{equation}

Combined with histogram loss, \textit{CC-WSI-Net} is optimized by the following: 
\begin{equation}
min_{G} ( \max_{D} \sum_{i=1}^2 \left( \log(D_i(X_s)) + \log(1 - D_i(G(X_h))) \right) + \lambda \times L_{\text{feat}} + L_{\text{DET}} + \text{W}_{h} \times \text{L}_{h})
\label{eq2}
\end{equation}

\section{Experiments and Results}
\subsection{Dataset}
\label{dataset}
In our study, we leverage two curated melanoma datasets: one for training and testing the model, and the other for subjective assessment of synthetic vs. real WSI quality by pathologists. Both datasets contain skin biopsy images stained with H\&E and Sox10. 

Dataset 1, the dataset used for training and testing, contains skin tissue samples from 15 cases randomly selected from a private dermatopathology lab (to be named in final paper). Slides were scanned in brightfield with a 20x Plan Apo objective using the NanoZoomer Digital Pathology whole slide scanning system (HT-9600;  Hamamatsu City, Japan).  Each case is categorized under the MPATH-Dx diagnostic category  system \cite{barnhill2023revision} and was sectioned into 4-6 slices, resulting in 75 slices at 20x magnification. Each slice was initially stained with H\&E and later carefully restained with Sox10, thus providing a matched pair of images for each slice: one H\&E-stained image and one Sox10-stained image. The slice images were then cropped into 27,538 patches of $256 \times 256$ pixels at 10x magnification. These patches were split into training, validation, and testing sets, with 15,567 patches used for training, 10,507 patches dedicated to testing (ensuring non-overlap with the training dataset), and 1,464 patches used for validation. We leverage the same preprocessing steps described in VSGD-Net \cite{liu2023vsgd} to register paired images and generate groundtruth melanocyte annotations for all the patches. This setup provides a comprehensive platform for evaluating melanocyte detection models across all MPATH-Dx diagnostic categories. 

Dataset 2, the dataset used for comparisons by pathologists, contains 25 new cases from a private dermatopathology lab (to be named in the final paper). Similar to the first dataset, each case is categorized under the MPATH-Dx diagnostic category system category to include the full spectrum of diagnostic classes of melanocytic skin lesions. Each case was sectioned into 6-12 slices, with each slice having both H\&E stained and Sox10-stained images. 

\subsection{WSI qualitative comparison}
\begin{figure}[ht]
    \centering
    \includegraphics[width=1\textwidth]{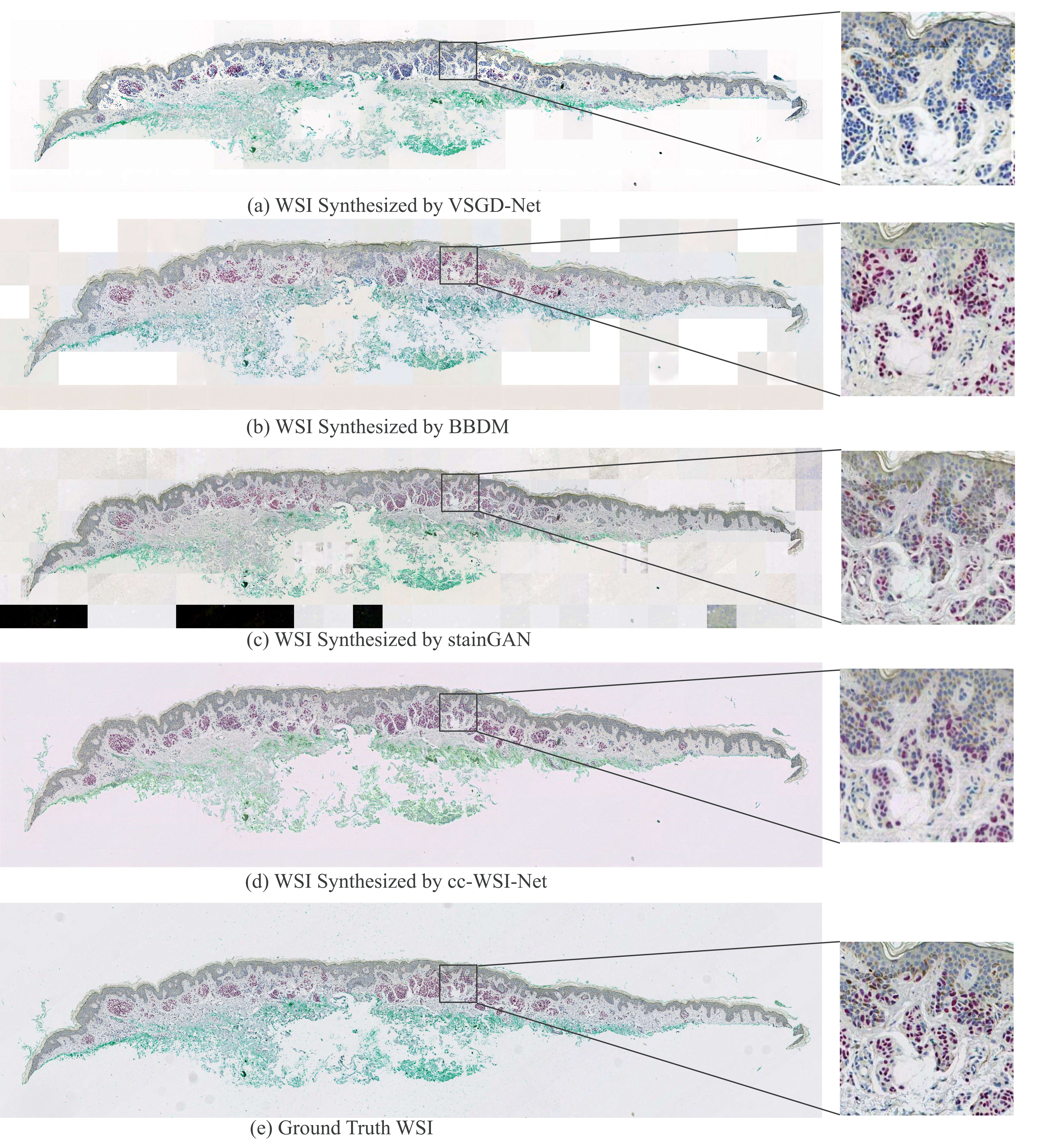} 
    \caption{Qualitative comparisons on synthesized WSI.}
    \label{WSIs_qualitative_comparison}
\end{figure}

As shown in Fig. \ref{WSIs_qualitative_comparison}, we provide qualitative comparisons on the WSIs synthesized by various models. The models used for comparison included the VSGD-Net, Brownian
Bridge Diffusion Model (BBDM) \cite{li2023bbdm}, and stainGAN \cite{shaban2019staingan}. Although all the synthesized WSIs are generated by stitching the patches, our \textit{CC-WSI-Net} successfully generated seamless WSIs without obvious inconsistencies. The WSIs synthesized by the VSGD-Net, BBDM and stainGAN all have content inconsistency problems and significant color inconsistency problems. Zoomed-in crops show the same position detail among all the WSIs. Since the content inconsistencies showed up in different positions, a zoomed-in crop of the WSI synthesized by BBDM  is provided to show the significant content inconsistency problems. Note that the zoomed-in crops of the VSGD-Net, BBDM and stainGAN have stitching artifacts on color, while the consistency modules in \textit{CC-WSI-Net} greatly benefit the synthesis quality.

\subsection{Melanocyte detection comparison}
To investigate the stain accuracy of the synthesized image and its potential to aid diagnosis, we evaluate the melanocyte detection performance using VSGD-Net, BBDM, and our \textit{CC-WSI-Net}. For VSGD-Net and \textit{CC-WSI-Net}, detected objects are directly obtained from the model itself. For BBDM, we apply the same ground truth extracting steps on the synthesized Sox10 images to obtain the predicted melanocytes. We report the average precision, recall and F1-score for Data Set 1 in Table \ref{Melanocytes detection performance comparisons}.
The precision, recall and F1-score of the \textit{CC-WSI-Net} framework are similar to that of VSGD-Net with higher precision and F1, and they are higher than the diffusion-based BBDM. Our \textit{CC-WSI-Net} framework can achieve comparable accuracy to the VSGD-Net, while synthesizing WSIs with better quality.  

\subsection{Training Details}
The model was trained on two NVIDIA GeForce GTX 1080 GPU with 11 GB of memory each. The model takes patches of size $256\times256$ as the input and generates patches of size $192\times192$.

\subsection{Ablation Study}

\subsubsection{Color Transfer}

\begin{figure}[ht]
    \centering
    \includegraphics[height=1.7cm]{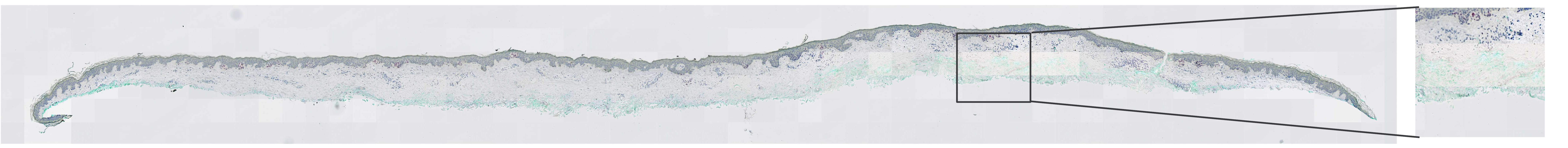} 
    \caption{WSI synthesized by the basic framework without the color consistency block and the color condition. The colors of the synthetic patches are only controlled by the histogram loss.}
    \label{fig:w_histo_loss_only}
\end{figure}

We ablated the components of the color consistency module. To verify the efficacy of the content consistency module, we trained a model using only the histogram loss. Fig. \ref{fig:w_histo_loss_only} demonstrates that, without the color consistency block and the color condition, although the color scheme is improved, the color inconsistency problem persists. Therefore, the color transferring block is needed in the framework for better control of the color scheme of the output.

\begin{table}[H]
\centering
\caption{Melanocytes detection performance comparisons}
\label{Melanocytes detection performance comparisons}
\begin{tabular}{|l|ccc|}
\hline
\textbf{Method} & \textbf{Precision} & \textbf{Recall} & \textbf{F1}    \\ \hline
VSGD-Net      & 0.671              & \textbf{0.641}  & 0.657          \\
stainGAN        & 0.373              & 0.080           & 0.132          \\
BBDM            & 0.555              & 0.418           & 0.477          \\ \hline
\textbf{Ours}   & \textbf{0.702}     & 0.630           & \textbf{0.664} \\ \hline
\end{tabular}
\end{table}

\subsubsection{Color condition}
The efficacy of using the 2D-histogram of a Sox10 WSI as the color condition was also tested. The image quality of the WSIs synthesized by CC-WSI-Net with different types of images as the color condition was evaluated. The types of color conditions included: 1) Sox10 WSI without the background (tissue masks were added to the Sox10 WSIs) and 2) ground truth Sox10 patches as the color condition. Since the WSIs synthesized by these conditions have similar qualitative result, we use Peak Signal-to-Noise Ratio (PSNR) and Root-Mean-Square Error (RMSE) to better compare the conditions used for synthesizing Sox10 WSI. PSNR measures image quality, where higher values indicate better quality of the synthetic patches. RMSE measures the similarity between the synthetic patches and the ground truth patches, with lower values indicating higher similarity. We report the average evaluation results among all test patches in Data Set 1 in Table \ref{tab:different color condition}. Both the WSI without background and the ground truth Sox10 patches have lower PSNR and higher RMSE compared to using the Sox10 WSI as the color condition, demonstrating the poor performance of these two color conditions. When H\&E WSIs without the corresponding ground truth Sox10 WSI were virtually stained using an arbitrary Sox10 WSI as the color condition, it became more challenging to control the color of the output. This difficulty arises because the histograms of WSIs without background and the histograms of patches differ significantly from the histogram of the WSI with background. Therefore, training the model with either the WSIs without background or patches results in poor image quality in the output.

\begin{table}[H]
\centering
\caption{Image quality of \textit{CC-WSI-Net} with a different color condition.}
\label{tab:different color condition}
\begin{tabular}{|l|ll|}
\hline
\textbf{Color Condition} & \textbf{PSNR}$\uparrow$ &  \textbf{RMSE}$\downarrow$ \\ \hline
WSI without background   & 19.550              & 8.386  \\
Ground Truth Sox10 Patch & 19.865              &8.251  \\ \hline
\textbf{WSI with background}   & \textbf{20.115}  & \textbf{7.978} \\ \hline
\end{tabular}
\end{table}

\section{Subjective Study with Pathologists}

\subsection{Study Design}

Since image quality metrics e.g. PSNR(Peak Signal-to-Noise Ratio), SSIM (Structural Similarity Index Measure), etc., fail to reflect the feasibility and effectiveness of synthetic Sox10 staining, a study is designed to combine quantitative accuracy assessment with rigorous qualitative analysis by experts, striking a balance between objective measurement and the subjective clinical relevance crucial for real-world applications. During the study, expert dermatopathologists were asked to review its performance compared to traditional Sox10 staining. Dataset 2, the set of 25 de-identified digital slides from skin biopsies categorized under MPATH-Dx classes 1 to 4\cite{barnhill2023revision} was used (Section \ref{dataset}). Importantly, these slides were independent of the training dataset used for \textit{CC-WSI-Net}. Each slide has a pair of synthetic Sox10 staining and traditional histochemical Sox10 staining for direct comparison.

\subsection{Pathologist Review Process}
Three board-certified pathologists (CH, SK, and DE) independently reviewed the digital slides using a Pathcore\textsuperscript{\textregistered}  interface. Each pathologist reviewed a total of 50 digital slides: 25 with traditional Sox10 staining and 25 with synthetic Sox10 staining. Each case included both H\&E and Sox10 stain images but omitted clinical history to ensure unbiased evaluations.
\par
To eliminate bias, the cases were divided into two blocks for presentation. In the first block, the 25 cases were presented in a random order, with each case assigned a specific staining method (either synthetic or traditional) through simple randomization. In the second block, the same 25 cases were presented again, but in a different random order, and each case was assigned the alternate staining method from the first block. This ensured that each pathologist reviewed every case twice, once with each staining method. Additionally, there were no demarcations to the pathologists that differentiated the two blocks during the review process. The sequence of the 50 slides was maintained consistently across all three pathologists to ensure uniformity in the evaluation.

\subsection{Evaluation Survey}
For each case, the pathologists were provided with an evaluation survey that included the following criteria:

\begin{enumerate}
    \item \textbf{Effectiveness of Sox10 Staining:} Rate how well the Sox10 staining made melanocytes more clearly visible and distinct from the surrounding tissue (1 = poor to 4 = perfect).
    \item \textbf{Image Quality:} Rate the overall quality of the whole slide image (1 = poor to 4 = perfect).
    \item \textbf{Staining Identification:} Indicate whether they believed the slide was synthesized, immuno-stained, or if they cannot tell.
\end{enumerate}

\subsection{Statistical Analysis}
The data collected from the evaluation surveys underwent basic descriptive statistical analysis. Mean ratings and percentages were calculated for image quality and staining effectiveness. 

\begin{table}
\centering
\caption{Review of 25 distinct cases, with one traditional Sox10 image and one synthetic Sox10 image each.  Three pathologists each reviewed the 50 Sox10 images, for a total of 150 reviews (N=75 reviews of traditional Sox10 images and N=75 reviews of synthetic Sox10 images).}
\label{tab: assessment: Review of 25 distinct cases}
\begin{tabular}{|l|rr|}
\hline
\textbf{Review Characteristics}          & \textbf{\begin{tabular}[c]{@{}r@{}}Traditional \\ Sox10 N (\%)\end{tabular}} & \textbf{\begin{tabular}[c]{@{}r@{}}Synthetic \\ Sox10 N (\%)\end{tabular}} \\ \hline
\textbf{Effectiveness of Sox10 Staining} &                                                                              & \multicolumn{1}{l|}{}                                                      \\
1 (poor)                                 & 13 (17\%)                                                                    & 6 (8\%)                                                                    \\
2                                        & 11 (15\%)                                                                    & 14 (19\%)                                                                  \\
3                                        & 29 (39\%)                                                                    & 32 (43\%)                                                                  \\
4 (perfect)                              & 22 (29\%)                                                                    & 23 (31\%)                                                                  \\ \hline
\textbf{Image Quality}                   & \multicolumn{1}{l}{}                                                         & \multicolumn{1}{l|}{}                                                      \\
1 (poor)                                 & 2 (3\%)                                                                      & 0 (0\%)                                                                    \\
2                                        & 1 (1\%)                                                                      & 0 (0\%)                                                                    \\
3                                        & 21 (28\%)                                                                    & 5 (7\%)                                                                    \\
4 (perfect)                              & 51 (68\%)                                                                    & 70 (93\%)                                                                  \\ \hline
\end{tabular}
\end{table}

\subsection{Evaluation Results}
The effectiveness and quality ratings for both traditional ground truth and synthesized Sox10 staining are shown in Table \ref{tab: assessment: Review of 25 distinct cases}. The synthetic Sox10 staining received higher mean ratings (standard deviation, sd) for both effectiveness and quality compared to traditional Sox10 staining. The mean (sd) rating for effectiveness for the synthetic images was 3.0 (0.9) and for traditional images was 2.8 (1.1).  The mean (sd) rating for image quality for synthetic images was 3.9 (0.3) and for traditional images was 3.6 (0.7).  The distribution of these ratings for effectiveness is further shown visually in Fig. 6.

The results of staining identification are shown in Table~\ref{tab: assessment: identifying the synthetic Sox10 images}. Pathologists reported that they could not distinguish the Sox10 staining method for 101 (67\%) of the images; when they placed a response attempting to guess if the image was traditional versus synthetic they were more likely to guess incorrectly. 

Overall, these findings suggest that the synthetic Sox10 WSIs generated by our method are indistinguishable from the true, traditional Sox10 WSIs. These evaluation results also indicate that CC-WSI-Net can produce high-quality virtual staining that maintains diagnostic relevance and effectiveness, comparable to traditional IHC methods.

Although these results are promising, the study has some limitations. The study was conducted in a controlled test environment with only three pathologists rather than in a clinical setting. While it enhances the reliability and repeatability of our tests, it does not fully replicate the complexities and unknown variables typical of clinical practice. Further studies are needed before clinical implementation. Additionally, technical elements such as slide preparation, potential variability in digitization techniques, and image processing methods need to be considered. For the traditional Sox10 stains, tissues were first stained with H\&E, the coverslip was removed, and then restained with Sox10. While this creates perfectly matched image pairs, it may introduce some imperfections in the Sox10 images. Despite standardized procedures, variability in tissue samples and staining remains possible. Future work should include a broader range of samples and settings to ensure the robustness and generalizability of our approach. Despite these limitations, this study has established a firm foundation for subsequent clinical validation and real-world application, promising to enhance diagnostic processes significantly.

\begin{figure}
    \centering
    \includegraphics[height=7cm]{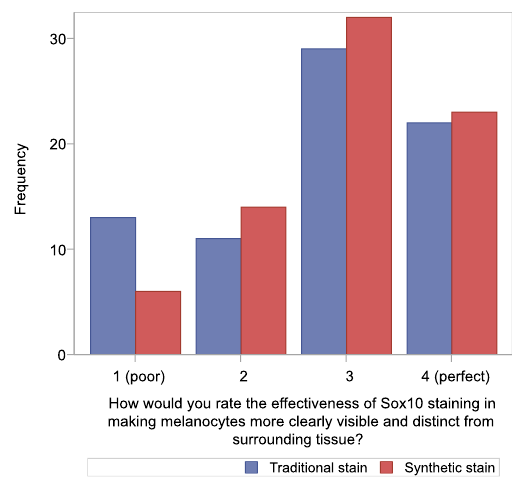} 
    \caption{Distribution of pathologist ratings of effectiveness of Sox10 staining for traditional and synthetic Sox10 images (N=150 reviews)}\label{fig:BarChart}
\end{figure}

\begin{table}[H]
\centering
\caption{Review of 25 distinct cases, with one traditional Sox10 image and one synthetic Sox10 image each. Three pathologists each reviewed the 50 Sox10 images, for a total of 150 reviews (N=75 reviews of traditional Sox10 images and N=75 reviews of synthetic Sox10 images).}
\label{tab: assessment: identifying the synthetic Sox10 images}
\begin{tabular}{|r|l|}
\hline
\multicolumn{1}{|c|}{\begin{tabular}[c]{@{}c@{}}Accuracy of pathologists in identifying Sox10\\ staining method\end{tabular}} & N (\%)                       \\ \hline
\multicolumn{1}{|l|}{\textbf{Incorrectly identified staining method}}                                                         & 37 (25\%)                    \\
Identified traditional when synthetic                                                                                         & 19 (13\%)                    \\
Identified synthetic when traditional                                                                                         & 18 (12\%)                    \\
\multicolumn{1}{|l|}{\textbf{Correctly identified staining method}}                                                           & 12 (8\%)                     \\
Correctly identified synthetic                                                                                                & \multicolumn{1}{c|}{8 (5\%)} \\
Correctly identified traditional                                                                                              & \multicolumn{1}{c|}{4 (3\%)} \\
\multicolumn{1}{|l|}{\textbf{Cannot tell}}                                                                                    & 101 (67\%)                   \\ \hline
\end{tabular}
\end{table}

\section{Conclusion}

In this study, we introduced \textit{CC-WSI-Net}, a framework for synthesizing virtually stained WSIs from H\&E images. This framework addresses color and content consistency issues between adjacent patches while ensuring stain accuracy. \textit{CC-WSI-Net} shows promise in synthesizing high-quality WSIs with different stains and can be incorporated into other models to solve color and content inconsistency problems.

\bibliographystyle{unsrt}  
\bibliography{references}  

\end{document}